\documentclass[12pt]{article}
\usepackage[dvips]{graphicx}
\usepackage{epsfig}
\usepackage{amsmath,amssymb,amsbsy}
\usepackage{amsfonts}
\usepackage{color}
\usepackage{soul}

\begin{document}

\begin{flushright}
LU TP 16-63\\
June 2017
\vskip0.5cm
\end{flushright}

\thispagestyle{empty}

\begin{center}
{\Large\bf{
The landscape of $W^{\pm}$ and $Z$ bosons 
\vskip 0.4cm 
produced in $pp$ collisions up to LHC energies}}
\vskip0.8cm
{\bf Eduardo Basso}
\vskip 0.2cm
Instituto de F\'isica, Universidade de S\~ao Paulo, Caixa Postal 66318, 05315-970 S\~ao Paulo, SP, Brazil
\vskip 0.4cm
{\bf Claude Bourrely}
\vskip 0.2cm
Aix Marseille Universit\'e, Universit\'e de Toulon, CNRS, CPT, UMR 7332 13288 Marseille, Cedex 09, France\\
%Aix Marseille Universit\'e, Universit\'e de Toulon, CNRS, Marseille, France\\
\vskip 0.4cm
{\bf Roman Pasechnik}
\vskip 0.2cm
Department of Theoretical Physics,\\
Lund University, SE 223-62 Lund, Sweden\\
\vskip 0.4cm
{\bf Jacques Soffer}
\vskip 0.2cm
Physics Department, Temple University,\\
1925 N, 12th Street, Philadelphia, PA 19122-1801, USA
\vskip 0.5cm

{\bf Abstract}
\end{center}

We consider a selection of recent experimental results on electroweak $W^{\pm},\,Z$ gauge boson production
in $pp$ collisions at BNL RHIC and CERN LHC energies in comparison to prediction of perturbative QCD calculations 
based on different sets of NLO parton distribution functions including the statistical PDF model known from fits 
to the DIS data. We show that the current statistical PDF parameterisation (fitted to the DIS data only) underestimates 
the LHC data on $W^{\pm},\,Z$ gauge boson production cross sections at the NLO by about 20\%. This suggests that 
there is a need to refit the parameters of the statistical PDF including the latest LHC data. 
%Besides, our analysis suggests some new tests for future data taking.
\vskip 0.1cm

%\noindent 

%\newpage 

%%%%%%%%%%%%%
\section{Introduction}
\label{sec:intro}
%%%%%%%%%%%%%
\setcounter{page}{1}

The ongoing measurements at particle colliders such as RHIC and the LHC continue precision tests of particle production 
mechanisms. In this respect, there is a growing demand for a better understanding of underlined QCD uncertainties, in particular,
related to modeling of parton density functions (PDFs), the key ingredients of QCD collinear factorisation. A major effort 
of the Particle Physics community over past decades has been directed towards constraining the QCD parton (quark and gluon) 
dynamics at various momentum scales connected, via DGLAP evolution, to the universal nonperturbative parton densities at some 
low scale $Q_0$. The latter are not fully predicted by the first QCD principles but are usually parametrized and extracted from the data.

In recent years, production of electroweak gauge bosons, both charged ($W^\pm$) and neutral ($Z^0,\,\gamma^*$), has attracted 
a lot of attention from theory and experiments as an important test of QCD (see e.g. Refs.~\cite{Mangano:2015ejw,Alioli:2016fum,
Basso:2015pba}). In particular, these processes are traditionally considered as an ideal tool for probing PDFs at various $x$ and $Q^2$ 
\cite{Hamberg:1990np,Martin:2009iq,Peng:2014hta}. For example, by controlling the c.m.s. energy $\sqrt{s}$, di-lepton rapidity $Y$ and 
invariant mass $M$ in the Drell-Yan (DY) process $pp\to (Z^0/\gamma^* \to l^+ l^-)+X$ one could access quark and gluon PDFs at both 
small and large $x_i=(M/\sqrt{s})e^{\pm Y}$ with $i=1,2$ denoting the incoming protons. While contributions from gluon and sea-quark PDFs 
to gauge boson production dominate presumably in kinematic regions of the LHC (except, probably, highly forward regions of the phase space), 
at lower energies of RHIC one expects an increased sensitivity to valence quark distributions. 

In our previous study \cite{bbps}, several most recent PDF parametrizations at the next-to-leading order (NLO) including the statistical PDF 
(known as NLO BS15) model \cite{bs15} were used for a description of the existing DY data for the normalised differential distributions available from 
Tevatron and LHC measurements. While a fairly good description of the DY data at high energies has been found for all the chosen PDF sets, at low energies 
the PDF models exhibit more substantial differences in shapes of the invariant mass and $x_F$ distributions. Provided that the BS15 model having 
much fewer free parameters results in as good data description as other popular models so it should be considered on the same footing as the current 
global PDF fits. In this paper, we extend our previous analysis \cite{bbps} to RHIC and LHCb kinematic regions incorporating also $W^\pm$ 
production observables and verify the predictions of the statistical PDF model at NLO accounting for resummation up to the Next-to-Leading 
Log (NLL) level against the corresponding experimental data.

The paper is organized as follows. In Section~\ref{sec:pdf-sets}, we review the main features of two sets of PDFs we have used for 
our calculations. In Section~\ref{sec:W-prod}, we consider a selection of recent $W^{\pm}$ production data sets available from 
STAR, CMS and LHCb and the DY-pair $Z/\gamma^*\to l\bar l$ production data -- from the STAR and LHCb measurements.
We make a comparison of theoretical predictions for the selected PDFs with these data. We give our summary and final remarks 
in Section~\ref{sec:summary}.

%%%%%%%%%%%%%%%%%%%%
\section{Selection of two key PDF sets}
\label{sec:pdf-sets}
%%%%%%%%%%%%%%%%%%%%

We will now summarize the essential properties of two sets of PDFs which will be tested in our further analysis of $W^{\pm}$ 
and DY-pair production in $pp$ collisions at various energies. 

The basic features of the statistical PDF approach alternative to canonical polynomial parametrizations which are
inspired by the Regge theory at small $x$ and by counting rules at large $x$ have been discussed in Ref.~\cite{bs15}.
In particular, (anti)quark distributions at the input scale $Q_0^2=1\,\mbox{GeV}^2$
\begin{eqnarray}
  && xq^h(x,Q^2_0) = \frac{A_{q}X^h_{0q}x^{b_q}}{\exp [(x-X^h_{0q})/\bar{x}]+1} +
  \frac{\tilde{A}_{q}x^{\tilde{b}_{q}}}{\exp(x/\bar{x})+1} \,, \label{eq1} \\
  && x\bar{q}^h(x,Q^2_0) = \frac{{\bar A_{q}}(X^{-h}_{0q})^{-1}x^{b_{\bar q}}}
  {\exp [(x+X^{-h}_{0q})/\bar{x}]+1} + \frac{\tilde{A}_{q}x^{\tilde{b}_{q}}}{\exp(x/\bar{x})+1} \,, \label{eq2}
\end{eqnarray}
are defined in terms a quasi Fermi-Dirac function (first terms) and a helicity independent diffractive 
component (second terms). Note, the latter does not enter the quark helicity $\Delta q$ and valence $q-\bar q$ 
distributions. The multipliers $X^{h}_{0q}$ and $(X^{-h}_{0q})^{-1}$ in the diffractive contributions have
been justified in the statistical approach to transverse momentum dependent PDFs in Ref.~\cite{bbs13}.
In Eqs.~(\ref{eq1}) and (\ref{eq2}) for a given quark $q$ with fixed helicity $h=\pm$ the parameters 
$\bar{x}$ and $X^{h}_{0q}$ play the role of universal temperature and thermodynamical potential 
encoding the main characteristics of the model (for antiquarks the sign of helicity and potentials is changed).
Remarkably, the statistical PDF approach enables to describe both upolarised observables and helicity asymmetries.
In what follows, however, only spin-independent observables are considered.

The statistical (anti)quark distributions contain in total eight parameters\footnote{It turns out that $X_{0u}^-$ 
and $X_{0d}^-$ were found almost identical.} for a given $q$,  namely, $\bar {A}_q$, $A_q$, $\tilde {A}_q$, 
$X^{\pm}_{0q}$, $\bar {b}_q$, $b_q$, and $\tilde {b}_q$. Then, the valence sum rule, \[\int (q(x) - \bar {q}(x))dx = N_q\,, 
\qquad N_q = 2, 1, 0 \] for  $u, d, s$ quarks, respectively, reduces the parameter space to seven free parameters.
The additional constraints apply for $q=\{u,d\}$ \cite{bbs1}
\begin{eqnarray}
\bar {A}_u = \bar {A}_d\,, \quad A_u=A_d\,,\quad b_u = b_d\,, \quad  \tilde {A}_u = \tilde {A}_d\,,\quad 
\tilde {b}_u = \tilde {b}_d \,, \quad \bar {b}_u = \bar {b}_d\,, 
\end{eqnarray}
thus, leading to eight free parameters in the sector of light quarks such that the diffractive contribution 
is flavor independent (for more details, see e.g.~Ref.~\cite{bbs-rev}). The expression for the statistical gluon 
PDF at $\mu=Q_0$ is inspired by the black-body spectrum and has a form of a quasi Bose-Einstein function
\begin{equation}
   xG(x,Q^2_0) = 
   \frac{A_Gx^{b_G}}{\exp(x/\bar{x})-1} \,, \label{eq3}
\end{equation}
where $A_G$ is found by the momentum sum rule such that $b_G$ is the only additional free parameter.

To conclude, the statistical PDF sets\footnote{In Ref.~\cite{bs15} we have also considered the helicity gluon distribution 
which is irrelevant in the present work.} contain {\it seventeen} free parameters in total. Besides the temperature $\bar x$ 
and the exponent of the gluon distribution $b_G$, the light $u,d$ and strange $s$ quark PDFs are constructed in terms of
{\it eight} and {\it seven} free parameters, respectively. These were fitted to a large set of accurate unpolarised and 
polarised DIS data {\it only} at the NLO QCD level and therefore are denoted as NLO BS15 from now on.

Let us discuss the second PDF set used in our calculations of the unpolarized cross sections below. Several versions were proposed by 
the CTEQ-TEA global analysis of QCD up to NNLO \cite{cteq} including data from HERA, Tevatron and LHC. For each flavor 
the Regge-motivated parametrization is of the form
\begin{eqnarray}
 x f_a(x,Q_0^2)  =  x^{a_1} (1-x)^{a_2} P_a(x) \,,
\label{xfa}
\end{eqnarray}
with a slowly-varying polynomial factor $P_a(x)$. In the CT14 model \cite{ct14} this factor for valence distributions 
is represented in terms of a linear combination of Bernstein polynomials
\begin{eqnarray}
&& P_{q_v} \, =  d_0  p_0(y)  +  d_1  p_1(y)  + 
 d_2  p_2(y)  +  d_3  p_3(y)  +  d_4  p_4(y) \,, \qquad y=\sqrt{x} \,, \nonumber \\
&& p_0(y)=(1-y)^4\,, ~p_1(y)=4 y (1 - y)^3\,, ~p_2(y)=6 y^2 (1 - y)^2\,, \nonumber \\ 
&& p_3(y) = 4  y^3  (1 - y)\,, ~p_4(y) = y^4 \,.
\end{eqnarray}
By fixing $d_1 = 1$, $d_3 = 1 + a_1/2$ and using the valence sum rule the number of free parameters for each flavor
is reduced to four. Thus, eight parameters fully determine the valence $u_v$ and $d_v$ distributions. Due to fewer constraints
on the gluon distribution, the CT14 gluon PDF is constructed in terms of a lower-order polynomial
\begin{eqnarray}
&& P_{g}(y')  =  g_0  \left[e_0  q_0(y')  +  e_1 q_1(y')  + q_2(y') \right] \,, \quad
q_0(y') = (1 - y')^2\,, \nonumber \\
&& q_1(y') = 2 y'  (1 - y')\,, \quad ~q_2(y') = {y'}^2 \,,
\end{eqnarray}
where $y'  =  1  -  (1-\sqrt{x})^2  =  2\sqrt{x} - x$. With an account for the momentum sum rule
the gluon PDF is determined by five free parameters in total. Fourth-order polynomials 
in the same variable $y'$ as for the gluon PDF were employed for building the sea 
$\bar{d}$ and $\bar{u}$ distributions assuming $\bar{u}(x)/\bar{d}(x) \to 1$ at $x \to 0$. 
Altogether, in the CT14 model the sectors of valence and sea quark PDFs contain {\it eight} and {\it thirteen} free parameters, 
respectively, while the gluon PDF contains {\it five} parameters amounting to {\it twenty six} fitting parameters in total. 
The (N)NLO QCD global fits in this model have been performed at $Q_0$=1.295 GeV.

In numerical analysis we employ the up-to-date computing tools available for high precision PDFs studies. 
All the numerical results presented below were obtained using the \texttt{DYRes} code \cite{dyres} which 
computes the DY observables up to NNLO performing the resummation of logarithms that become large when 
the vector boson transverse momentum is much smaller than the boson mass. 
Such a resummed result is matched order by order, up to $\mathcal{O}(\alpha_s^2)$, 
with the fixed-order result at large transverse momenta which is obtained within the dipole formalism 
\cite{Catani:2000vq} and is implemented in the \texttt{MCFM} code \cite{Campbell:1999ah,mcfm}. 
The divergences in the fixed-order calculation at small $p_T$ are subtracted resulting in a modification 
of the \texttt{DYNNLO} program \cite{dynnlo} which was used in our previous DY analysis \cite{bbps}. 

Although the resummation may not improve the results at larger values of transverse momentum being 
probed in the present study, its results have shown to agree with those from fixed-order calculations \cite{dyres}. 
Since both the resummed and the fixed order calculation are expected to agree at large $p_T$, we will 
proceed using \texttt{DYRes} at the NLL level throughout all the predictions below unless there is a notable 
disagreement between \texttt{DYRes} and the fixed-order results provided by \texttt{DYNNLO}. We indeed 
observed an overall agreement between both frameworks for most of the results presented bellow, with an exception 
that \texttt{DYNNLO} results on the $W^+/W^-$ ratio at NLO appear to describe the current LHCb data 
noticeably better than the \texttt{DYRes} ones.

%%%%%%%%%%%%%%%%%%%%%%%%%%%%%%%%%%%%
\section{Results for $W^\pm$ and $Z/\gamma^*$ observables}
\label{sec:W-prod}
%%%%%%%%%%%%%%%%%%%%%%%%%%%%%%%%%%%%

Our previous analysis of recent $Z$-boson production observables measured at different energies 
including recent LHC data in Ref.~\cite{bbps} (and normalised to the total cross section) has shown 
an overall very good consistency of the BS15 PDF model with the data. Does the same situation 
persist for $W^\pm$ observables?

Note, in what follows we use the BS15 model that was fitted to the DIS data only and apply it in analysis
of the data obtained at energies as high as the LHC ones. This step is needed to estimate the underlined
theoretical uncertainties and to justify whether or not a global fit (including the LHC data) is needed for
the statistical PDF parameterisation. In all the figures below, we have used \texttt{DYRES} to the NLL accuracy 
which is sufficient to see an overall consistency with the data. Clearly, when global fits with BS model at 
the NLO are performed in the future, one could go higher in the resummation accuracy if needed.

\begin{figure}[h!]
\begin{center}
\includegraphics[width=12.0cm]{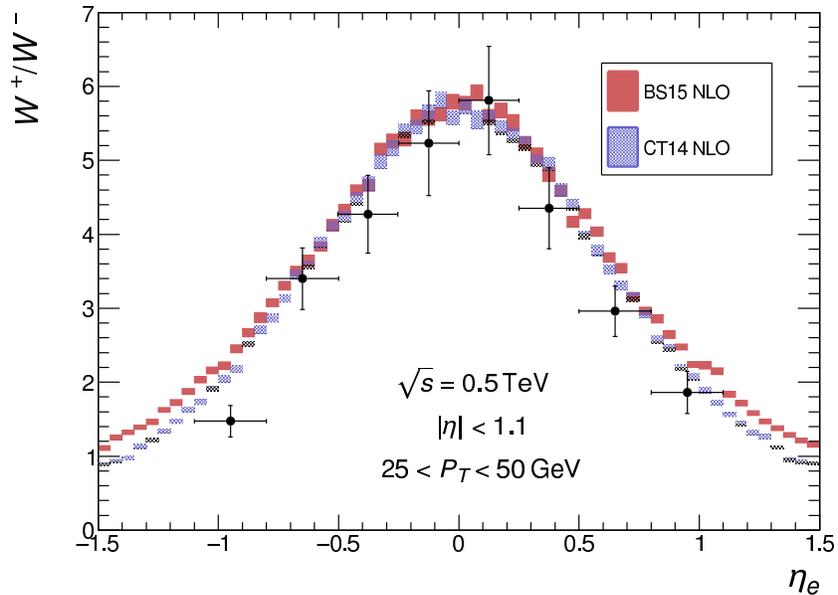}
\caption[*]{\baselineskip 1pt
The ratio the $W^+$ and $W^-$ differential cross sections at $\sqrt{s} = 500\,\mbox{GeV}$, versus the pseudo-rapidity 
of the charged lepton $\eta_e$, obtained with the BS15 and CT14 PDF models and compared to the preliminary data 
from the STAR experiment~\cite{mp}.}
\label{WpWmSTAR}
\end{center}
\end{figure}

\begin{figure}[h!]
\begin{center}
\includegraphics[width=6.75cm]{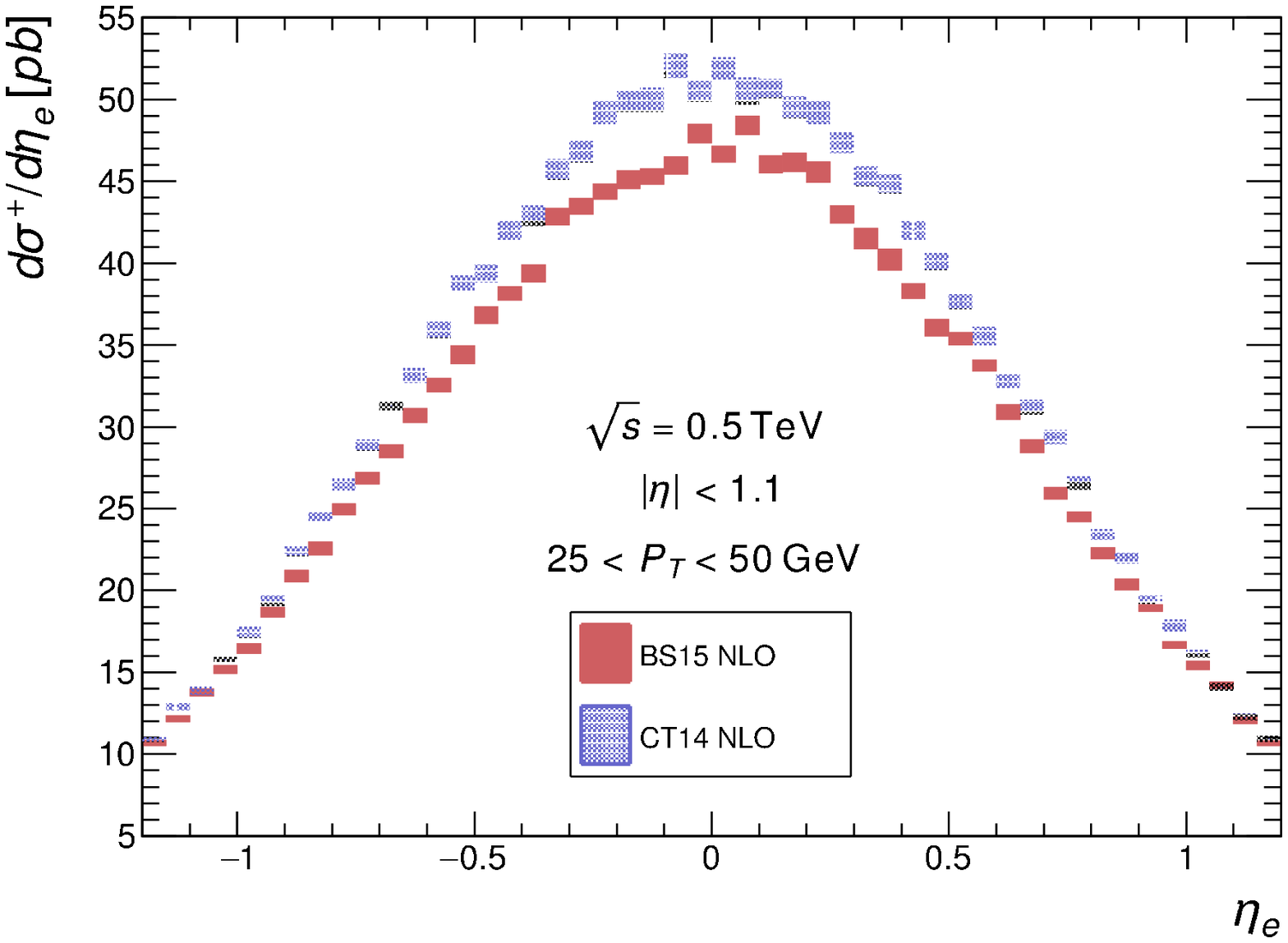}
\includegraphics[width=6.75cm]{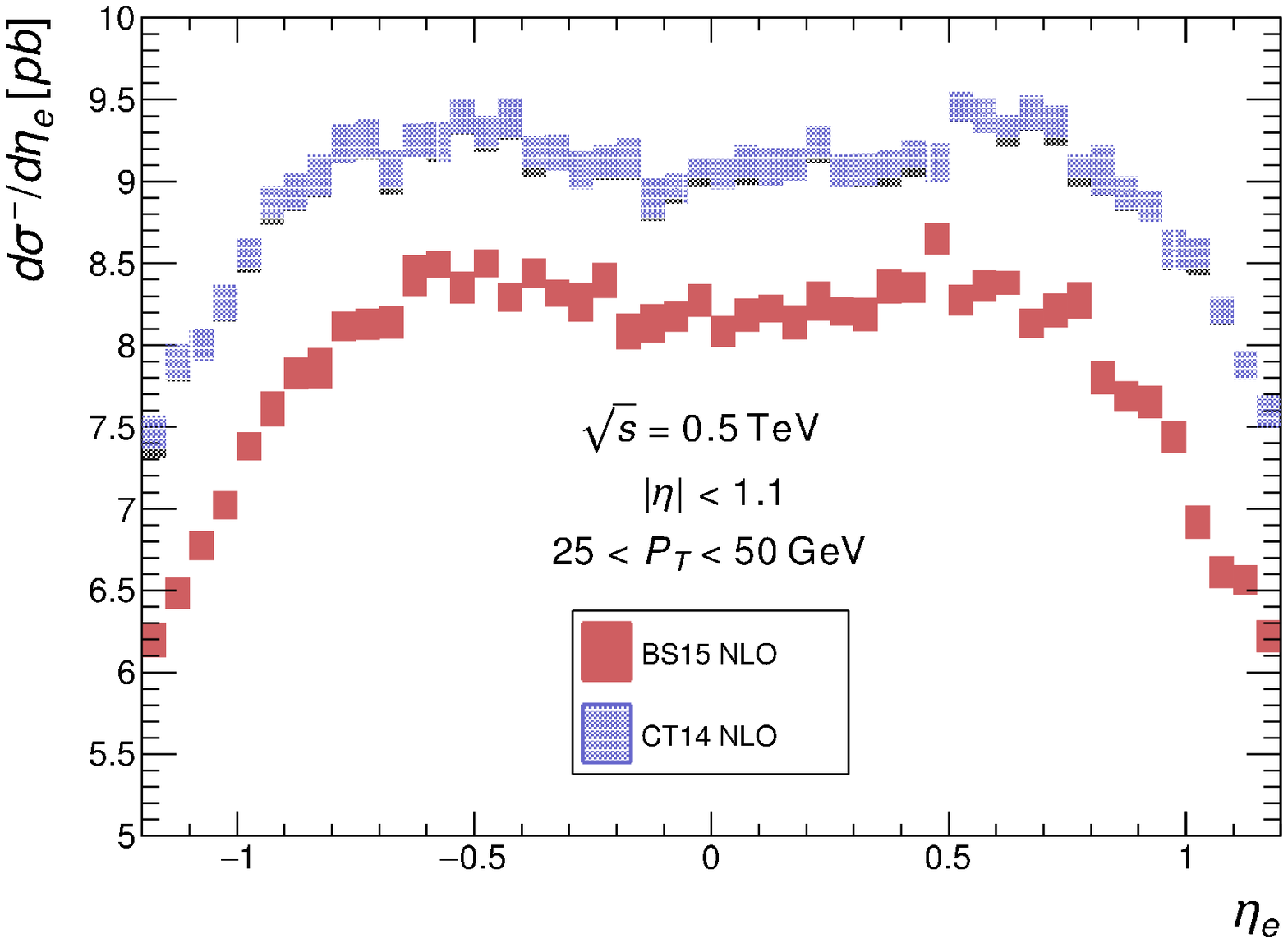}
\caption[*]{\baselineskip 1pt
The differential cross sections for $W^+$, $d \sigma^{+}(\eta_e)/d \eta_e$ (left), and $W^-$, $d \sigma^{-}(\eta_e)/d \eta_e$ (right),
production at $\sqrt{s} = 500\,\mbox{GeV}$, versus pseudo-rapidity of the charged lepton $\eta_e$, with the cuts corresponding 
to the acceptance of the STAR detector, as predicted by the BS15 and CT14 PDF models.}
\label{WpandWmSTAR}
\end{center}
\end{figure}

Consider first, as displayed in Fig.~\ref{WpWmSTAR}, a comparison of preliminary STAR data~\cite{mp} on 
the $W^+$ to $W^-$ differential cross section ratio 
\[
W^+/W^- \equiv \frac{d \sigma^{+}(\eta_e)/d \eta_e}{d \sigma^{-}(\eta_e)/d \eta_e}
\] 
with theoretical predictions based on the BS15 and CT14 PDF models. It is clear that both PDFs are consistent with the trend of the data given its poor uncertainty, 
but the planned 2017 run at BNL RHIC will be able to increase substantially the precision on this measurement. We observe that both predictions are almost sitting 
on top of each other, but they would be more distinguishable for larger $\eta_e$. However, this region is not accessible to STAR and the only way to find out which 
one agrees best with the data is to measure the individual cross sections near $\eta_e = 0$. This is indeed what one sees clearly in Fig.~\ref{WpandWmSTAR}, 
where the BS15 result remains below the CT14 one in both $W^+$ and $W^-$ cases. We urge the STAR Collaboration to perform this important test.
%%%%%%%%%%%%%%%%%%%%%%%%%%%%%%%%%%%%
%%%%%%%%%%%%%%%%%%%%%%%%%%%%%%%%%%%%
%%%%%%%%%%%%%%%%%%%%%%%%%%%%%%%%%%%%
%%%%%%%%%%%%%%%%%%%%%%%%%%%%%%%%%%%%

\begin{figure}[h!]
\begin{center}
\includegraphics[width=6.75cm]{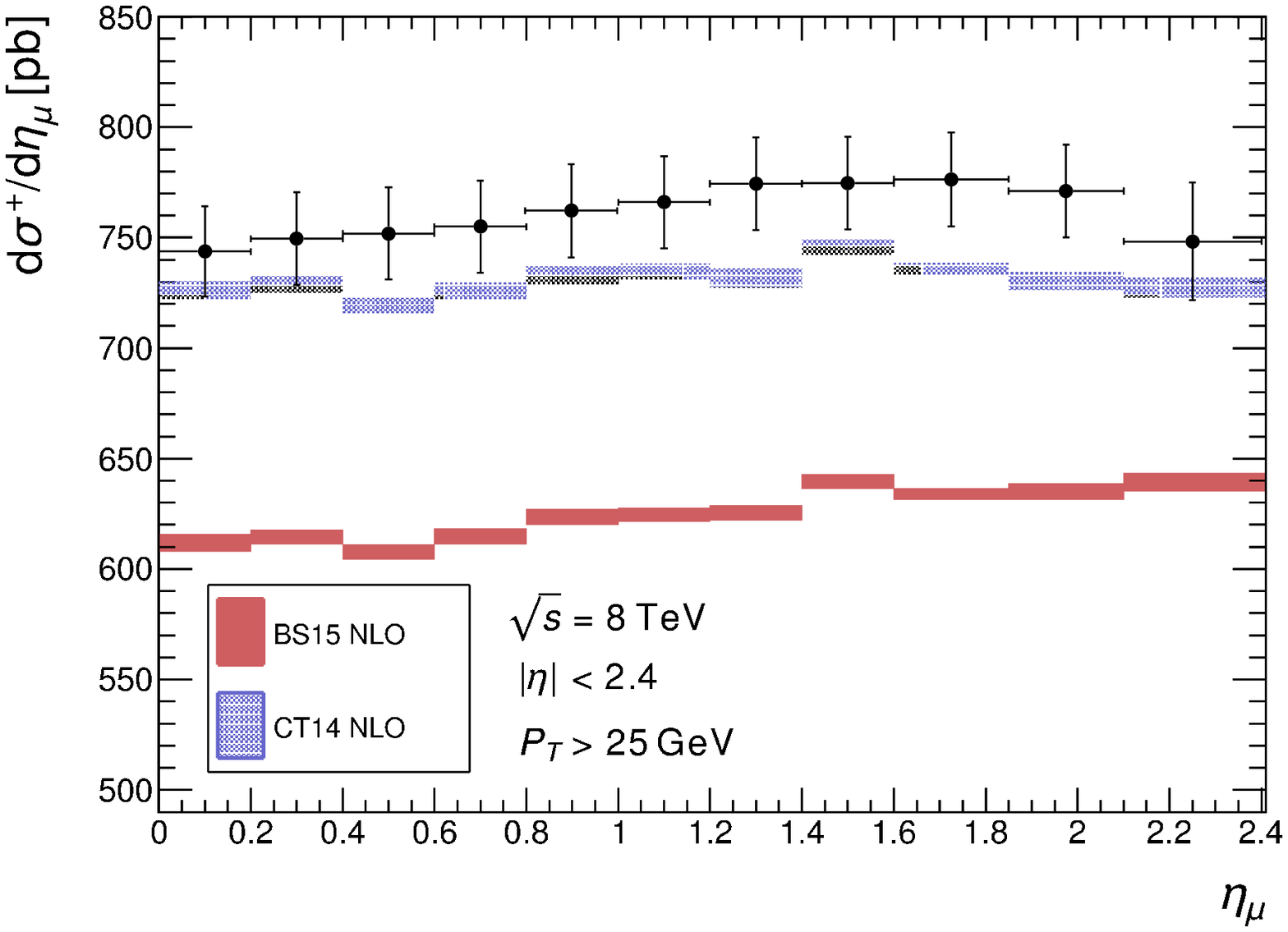}
\includegraphics[width=6.75cm]{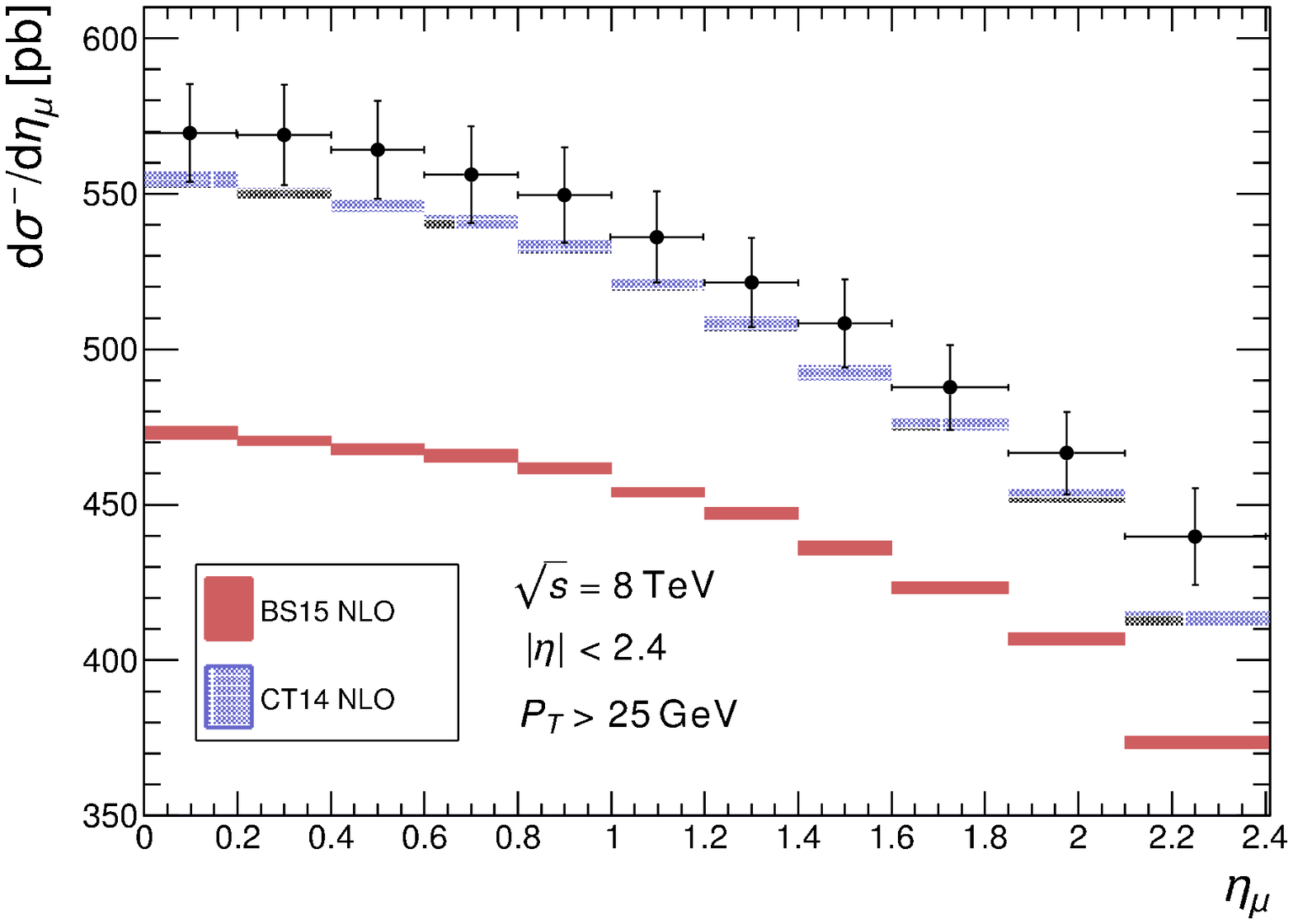}
\caption[*]{\baselineskip 1pt
The differential cross sections for $W^+$, $d \sigma^{+}(\eta_{\mu})/d \eta_{\mu}$ (left), and $W^-$, $d \sigma^{-}(\eta_{\mu})/d \eta_{\mu}$ (right),
production at $\sqrt{s} = 8\,\mbox{TeV}$, versus pseudo-rapidity of the charged lepton $\eta_{\mu}$, with the cuts corresponding 
to the acceptance of the CMS detector, as predicted by the BS15 and CT14 PDF models, in comparison to the data by CMS Collaboration~\cite{cms16}.}
\label{WpandWmCMS8Kfactor}
\end{center}
\end{figure}

The individual $W^\pm$ differential cross sections have been measured at the LHC at $\sqrt{s} = 8\,\mbox{TeV}$ by the CMS Collaboration~\cite{cms16} and 
are shown in Fig.~\ref{WpandWmCMS8Kfactor} together with the corresponding theoretical predictions using the BS15 and CT14 PDF models. While the general 
trends for the $W^+$ cross section are similar to that of the data for both BS15 and CT14 models, the shape of the BS15 prediction for the $W^-$ cross section 
somewhat deviates from the data in the high $\eta_{\mu}$ region. As we shall see below, this implies a prediction for the charge asymmetry that misses the data 
by a few percent. Unlike the neutral-current $Z$-boson observables discussed earlier in Ref.~\cite{bbps}, the CT14 and especially BS15 predictions fail to describe 
the charged-current precision data. The BS15 model prediction is off by about 20\% while the CT14 is doing much better and is only 3\% off the data in overall normalisation. 
These potentially indicate a strong need in making a global fit of the BS15 NLO model parameters including the LHC data. This stricking issue with normalisation was 
not seen earlier in the $Z^0$-boson production observables that were always normalised to the total cross section and strongly suggests future studies in this direction.
%%%%%%%%%%%%%%%%%%
%%%%%%%%%%%%%%%%%%
%%%%%%%%%%%%%%%%%%
\begin{figure}[h!]
\begin{center}
\includegraphics[width=6.75cm]{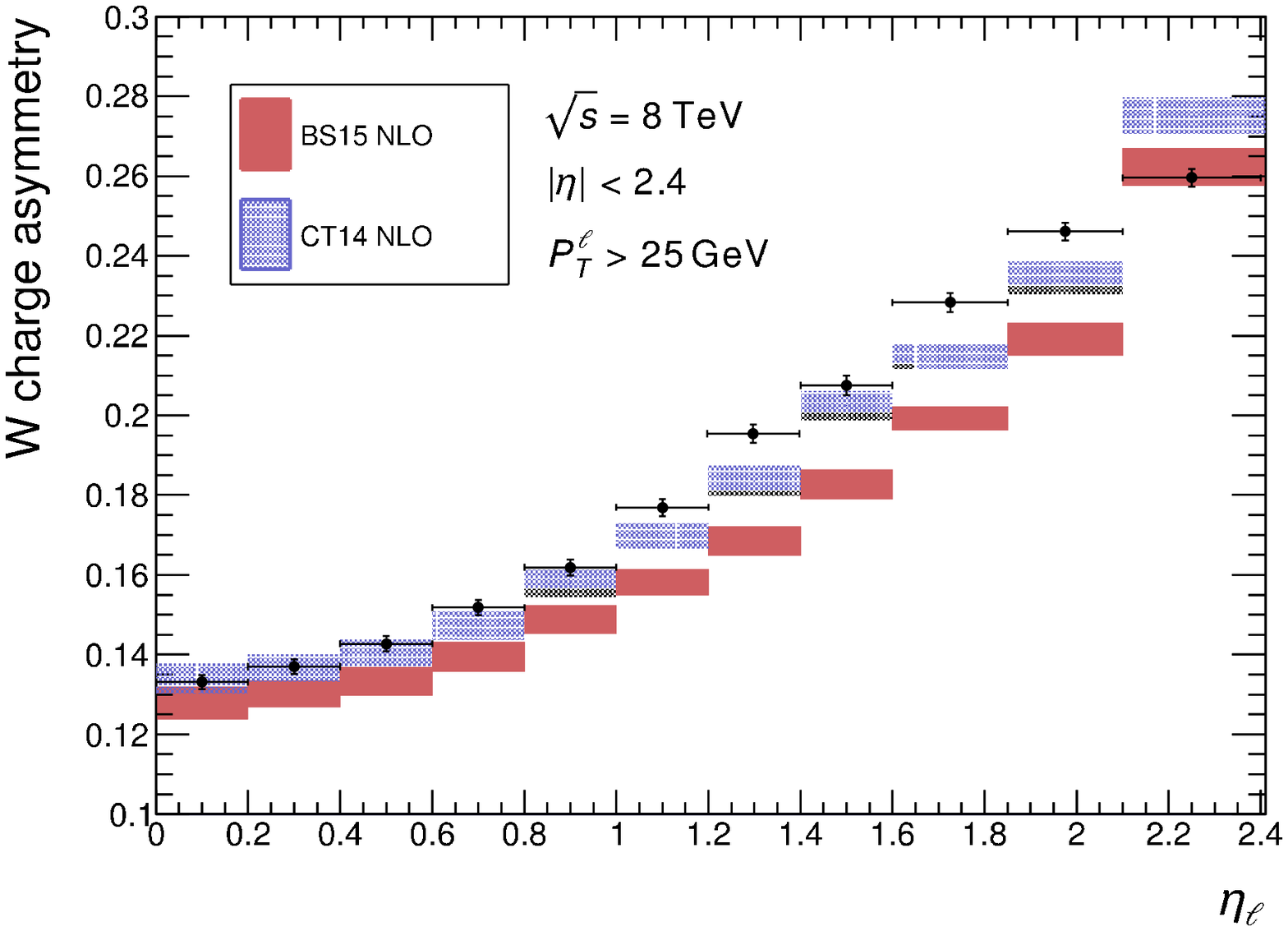}
\includegraphics[width=6.75cm]{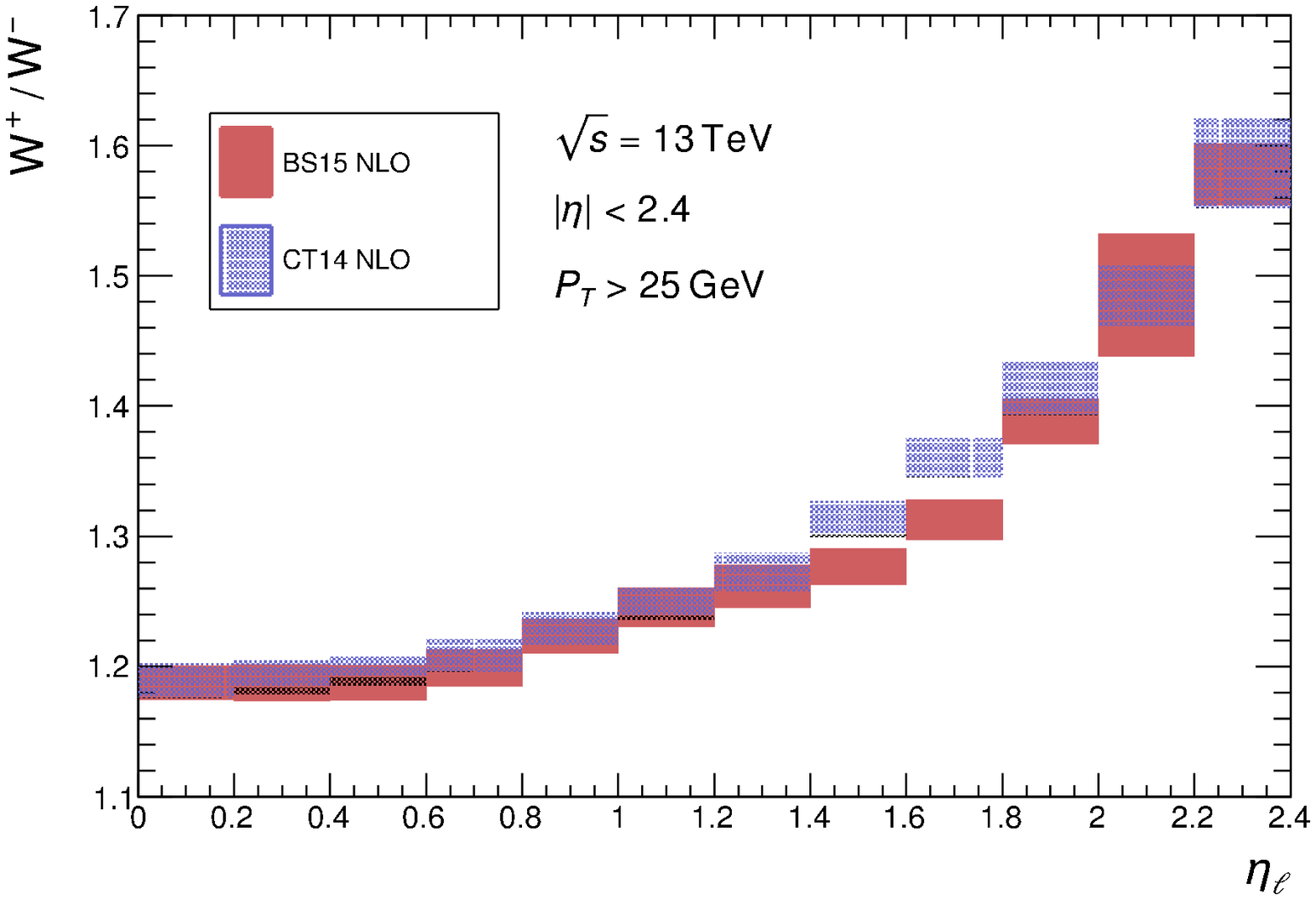}
\caption[*]{\baselineskip 1pt
The $W^\pm$ charge asymmetry at $\sqrt{s} = 8\,\mbox{TeV}$, versus pseudo-rapidity of the charged lepton $\eta_{\mu}$, with 
the cuts corresponding to the acceptance of the CMS detector, as predicted by the BS15 and CT14 PDF models, is shown in 
the left panel in comparison to the data by CMS Collaboration~\cite{cms16}. The prediction for $W^+/W^-$ ratio based on 
BS15 and CT14 PDF models at $\sqrt{s} = 13\,\mbox{TeV}$ are shown in the right panel.
}
\label{WasymmCMS}
\end{center}
\end{figure}
%%%%%%%%%%%%%%%%%%

\begin{figure}[h!]
\begin{center}
\includegraphics[width=12.0cm]{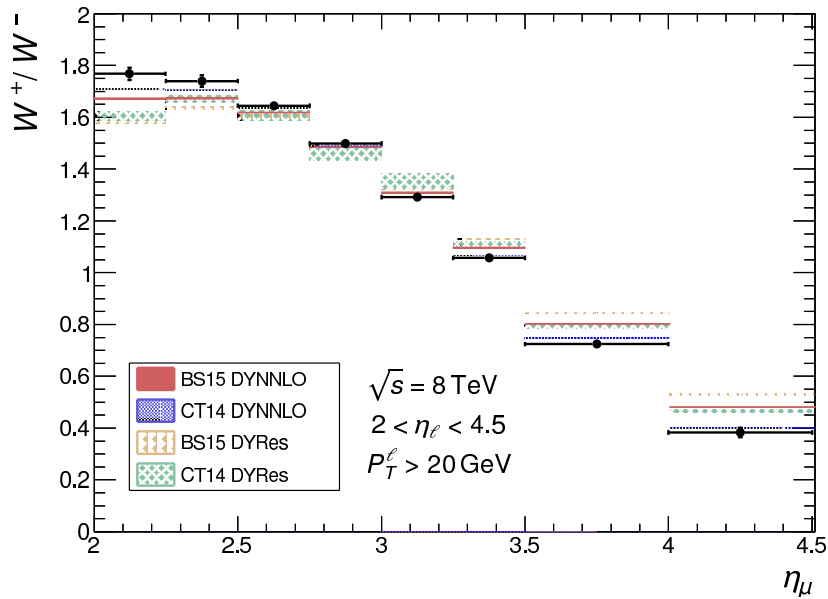}
\caption[*]{\baselineskip 1pt
The $W^+/W^-$ ratio based on BS15 and CT14 PDF models at $\sqrt{s} = 8\,\mbox{TeV}$ in comparison 
to the data from the LHCb Collaboration~\cite{LHCbZ8}. The fixed-order \texttt{DYNNLO} (NLO) results 
are compared to the resummed \texttt{DYRES} (NLO+NLL) results.
}
\label{WpWmLHCb}
\end{center}
\end{figure}

Another way to present the data is the charge asymmetry defined as 
\[
\frac{d \sigma^{+}(\eta_{\mu})/d \eta_{\mu} - d \sigma^{-}(\eta_{\mu})/d \eta_{\mu}}
{d \sigma^{+}(\eta_{\mu})/d \eta_{\mu} + d \sigma^{-}(\eta_{\mu})/d \eta_{\mu}}  \,,
\] 
which is related to $W^+/W^-$ ratio considered above. This is shown in Fig.~\ref{WasymmCMS} (left), but the disagreement of the BS15 and CT14 
predictions with the data persists for a limited $\eta$ region. The shape for BS15 model in the case of $W^-$, as seen in Fig.~\ref{WpandWmCMS8Kfactor}, 
is the main source of deviation in the asymmetry. In Fig.~\ref{WasymmCMS} (right) we present the predictions for $W^+/W^-$ ratio at 
$\sqrt{s} = 13\,\mbox{TeV}$. This ratio was also measured at $\sqrt{s} = 8\,\mbox{TeV}$ in a large pseudo-rapidity region by the LHCb 
Collaboration~\cite{LHCbZ8} (see Fig.~\ref{WpWmLHCb}). In order to compare the fixed-order (NLO) vs resummed (NLO+NLL) results with 
the data in Fig.~\ref{WpWmLHCb} we show both \texttt{DYNNLO} and \texttt{DYRES} results. We notice that \texttt{DYNNLO} works somewhat 
better against the data than \texttt{DYRES} in low $\eta$ region for both PDF parameterisations. Even in the case of the fixed-order \texttt{DYNNLO} 
analysis, one again notices a larger deviation from data for BS15, which reinforces the importance of a detailed analysis of the statistical PDF model, 
including global fits to the LHC data, to investigate the reasons for the $W^-$ deviation in shape from data, and also to correct for the overall 
normalization when considering LHC data. Such an analysis is beyond the scope of the preset work, but should be carried out in the future 
to improve the reliability of the statistical model.
%%%%%%%%%%%%%%%%%%
%%%%%%%%%%%%%%%%%%

\begin{figure}[h!]
\begin{center}
\includegraphics[width=12.0cm]{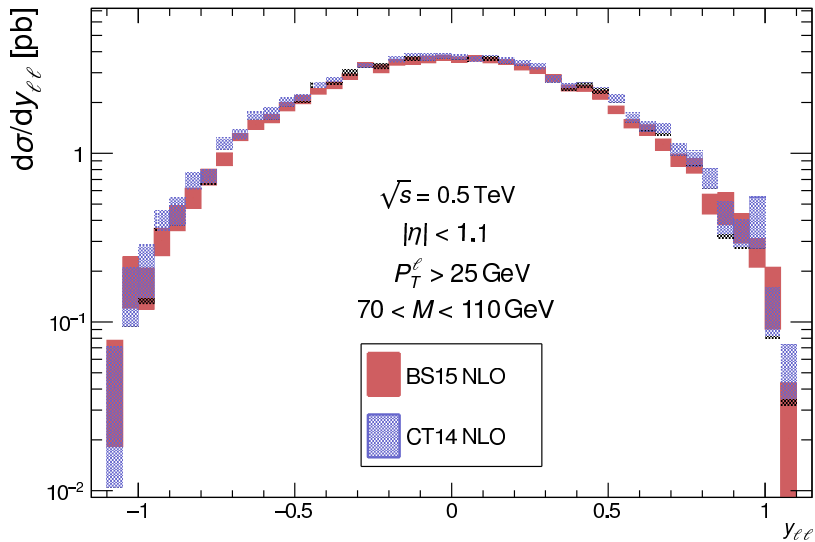}
\caption[*]{\baselineskip 1pt
The differential cross section for DY-pair $Z/\gamma^* \to l\bar l$ production, versus dilepton rapidity, 
with the cuts corresponding to the acceptance of the STAR detector. The predictions use the BS15 and CT14 PDF 
models, for comparison.}
\label{ZSTAR}
\end{center}
\end{figure}

\begin{figure}[h!]
\begin{center}
\includegraphics[width=6.83cm,height=4.85cm]{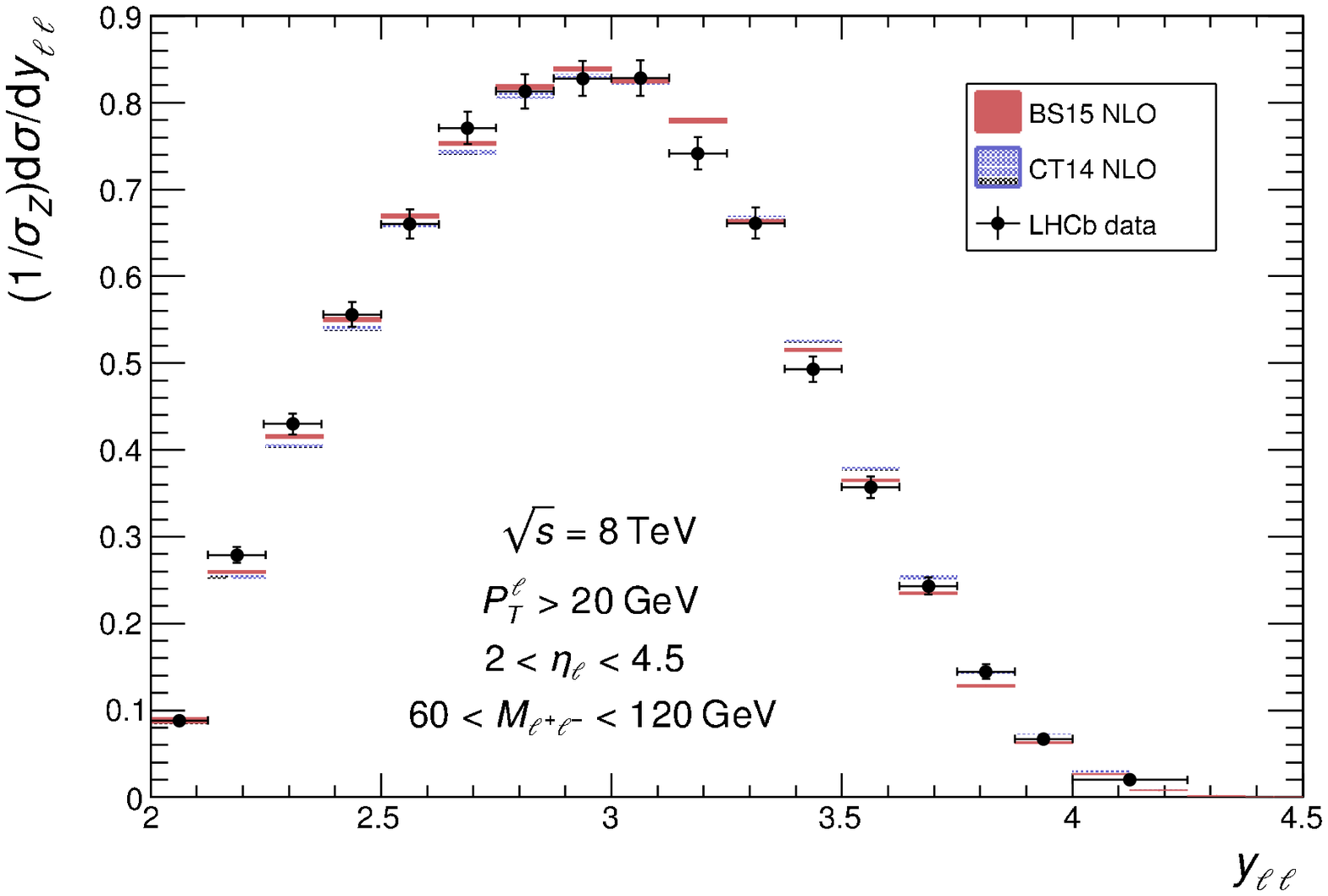}
\includegraphics[width=6.75cm]{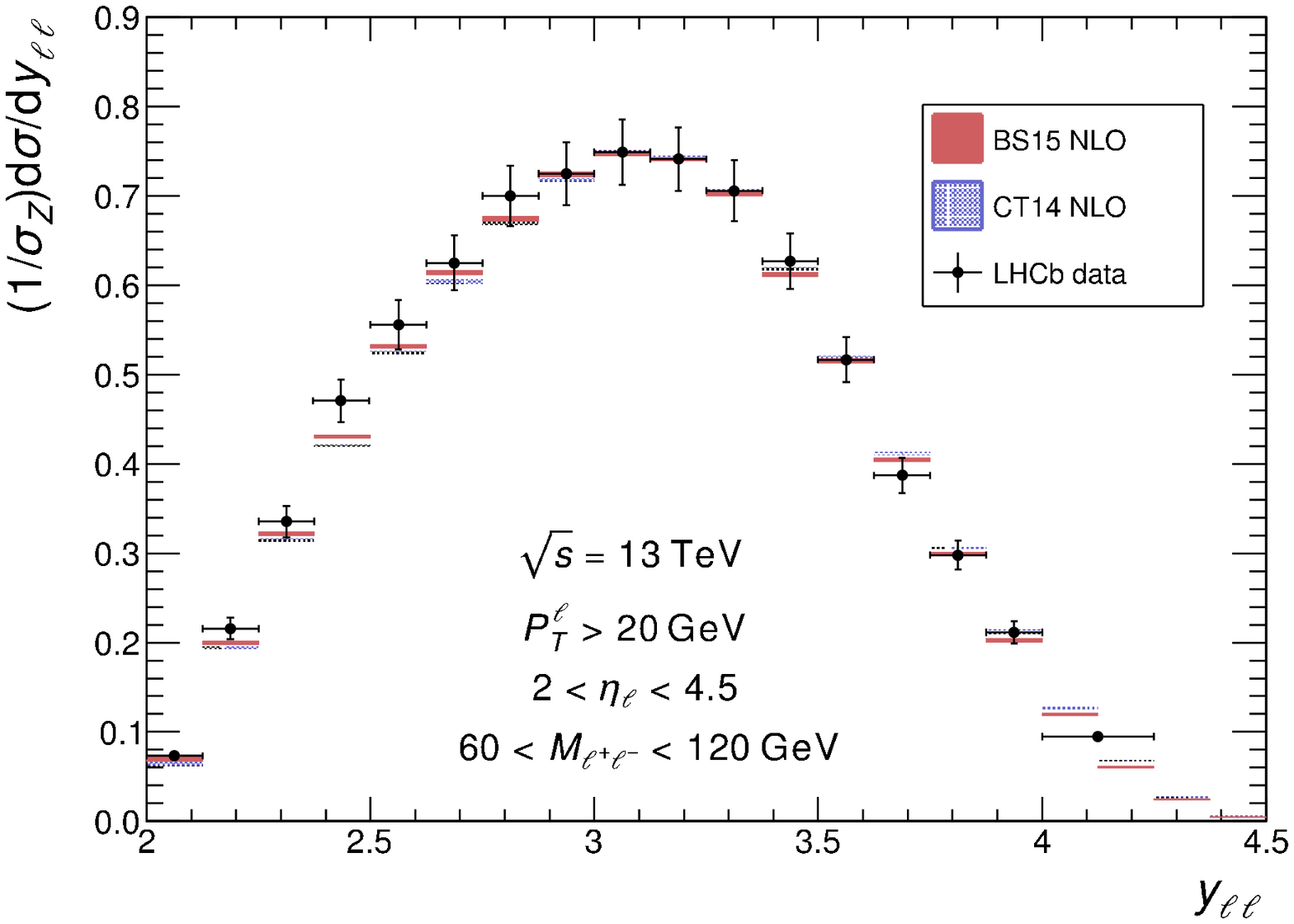}
\caption[*]{\baselineskip 1pt
The normalized differential cross section for the forward DY-pair $Z/\gamma^* \to l\bar l$ production as a function of dilepton rapidity against the data 
by the LHCb Collaboration, at $\sqrt{s} = 8\,\mbox{TeV}$ (left) \cite{LHCbZ8} and at $\sqrt{s} = 13\,\mbox{TeV}$ (right) \cite{LHCbZ13}. 
The experimental data are compared to the predictions using the BS15 and CT14 PDF models.}
\label{ZLHCb}
\end{center}
\end{figure}

\begin{figure}[h!]
\begin{center}
\includegraphics[width=12.0cm]{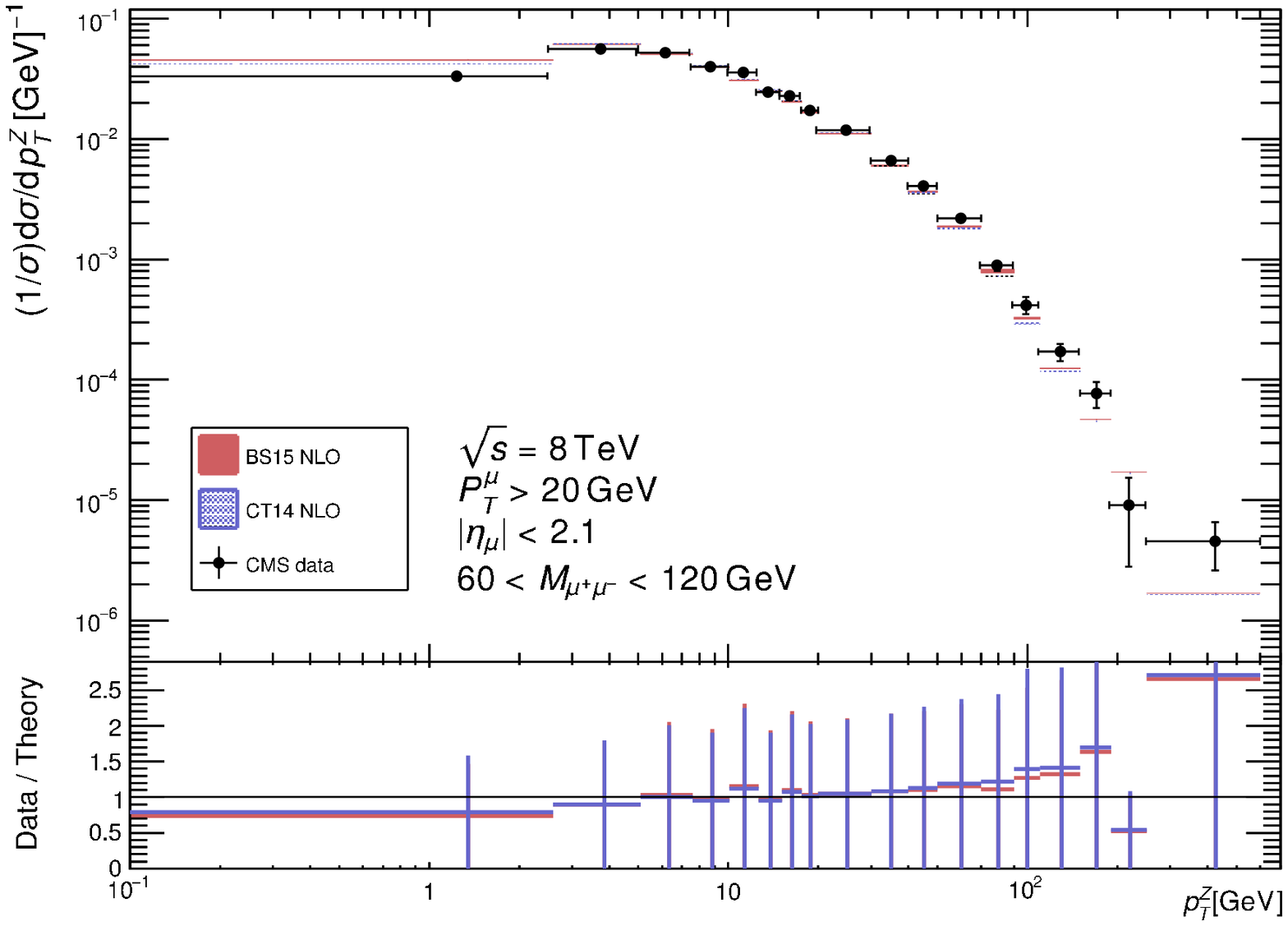}
\caption[*]{\baselineskip 1pt
The differential cross section for DY-pair $Z/\gamma^* \to \mu^+\mu^-$ production, versus dilepton transverse momentum, against the data 
by the CMS Collaboration \cite{Khachatryan:2016nbe}, with the cuts corresponding to the acceptance of the CMS detector. The predictions use 
the BS15 and CT14 PDF models, for comparison.}
\label{ZCMS}
\end{center}
\end{figure}

Several important aspects of the available data on $Z/\gamma^*$ production have already been studied in our earlier paper~\cite{bbps}.
In addition, in Fig.~\ref{ZSTAR} we show the predictions using the BS15 and CT14 PDFs for the differential cross section for $Z/\gamma^*$ production, 
versus the dilepton rapidity, in view of a future data taking from STAR at BNL RHIC. Once again it is not possible to distinguish them in this limited 
kinematic region. Indeed, an insufficient precision of the current experimental data makes it difficult to verify the PDF models so further improvements
toward a reduction of experimental uncertainties are needed. In Fig.~\ref{ZLHCb}, we display the normalized differential cross sections for the forward 
$Z/\gamma^*$ production from the LHCb experiment at two different energies ($\sqrt{s} = 8\,\mbox{TeV}$ \cite{LHCbZ8} and $\sqrt{s} = 13\,\mbox{TeV}$ 
\cite{LHCbZ13}) which turn out to be in a very good agreement with the predictions using the BS15 and CT14 PDF models. Finally, a similar situation holds 
for BS15 and CT14 predictions for the dilepton transverse momentum distribution against most recent CMS data at $\sqrt{s}=8$ TeV as shown in Fig.~\ref{ZCMS}.
%%%%%%%%%%%%%%%%%%
%%%%%%%%%%%%%%%%%%
%%%%%%%%%%%%%%%%%%
%%%%%%%%%%%%%%%%%%
%%%%%%%%%%%%%%%%%%
%%%%%%%%%%%%%%%%%%

%%%%%%%%%%%%%%%%%%
\section{Summary}
\label{sec:summary}
%%%%%%%%%%%%%%%%%%

As a natural continuation of our previous study on neutral-current $Z$-boson production observables \cite{bbps}, the main purpose of this analysis is to further 
test the statistical BS15 NLO model for parton density functions fitted only to DIS data versus the most recent RHIC and LHC data on charged-current $W^\pm$ bosons 
production. 

In this paper, we have studied the $W^\pm$ charge asymmetry and the differential (in lepton pseudorapidity) $W^\pm$ cross sections, as well as the differential 
Drell-Yan (DY) pair $Z/\gamma^* \to l\bar l$ production cross sections, at RHIC ($\sqrt{s} = 500\,\mbox{GeV}$) and LHC ($\sqrt{s} = 8,13\,\mbox{TeV}$) energies. 
In our analysis, we have used two distinct PDF sets at the NLL+NLO accuracy -- the statistical BS15 model and the CTEQ CT14 parametrization. The corresponding 
predictions obtained by using the \texttt{DYRes} package are compared to the most recent data sets available from RHIC and the LHC. 

The analysis of neutral-current (DY) observables (normalised to the total cross section) results in a fairly good description of the latest data, in full consistency with 
our previous analysis in Ref.~\cite{bbps}. Now, we have looked into such observables as the dilepton rapidity distributions (normalised to the total DY $Z/\gamma^*$ 
cross section) at $\sqrt{s}=8$ and 13 TeV in the LHCb kinematic regions, as well as the dilepton transverse momentum distribution at $\sqrt{s}=8$ TeV. No noticeable 
deviations in shapes of the BS15 and CT14 predictions at NLO+NLL versus data have been observed. The normalisation of the DY differential cross sections is not reproduced by 
both PDF models although CT14 NLO predictions are closer to the data than the BS15 ones, in particular, due to the fact the DY LHC data were not included 
into the BS15 fits.

Both BS15 and CT14 models work quite well also against the most recent data from RHIC on the $W^+/W^-$ ratio in the electron pseudorapidity region 
$-1.5<\eta_e<1.5$ at $\sqrt{s}=500$ GeV. For a more definite conclusion, one should have data on the individual $W^\pm$ distributions since 
the corresponding BS15 and CT14 predictions for the $W^+$ and $W^-$ distributions differ in overall normalisation by a few percent.

Notably, the BS15 model starts to exhibit larger discrepancies with respect to the $W^\pm$ production data at the LHC energies. In particular, its predictions 
for both $W^+$ and $W^-$ total cross sections are off by about twenty percent with respect to the CMS data at $\sqrt{s}=8$ TeV. While the shape of $W^+$ 
pseudorapidity distribution, $d\sigma^+/d\eta_\mu$, marginally reproduces that of the data, the shape of the $W^-$ pseudorapidity distribution, $d\sigma^-/d\eta_\mu$, 
is flatter and somewhat deviates from that of the data leading to a few-percent discrepancy at $\eta_\mu\simeq 1.5-2.0$. The latter discrepancy then translates 
into the corresponding deviations of the $W$ charge asymmetry at mid-pseudorapidities as compared to the CMS data. 

Interestingly enough, at forward pseudorapidities in the LHCb kinematic region, the shape of the lepton pseudorapidity distributions are in an overall consistency 
with our above conclusions at mid and central pseudorapidities. This is suggested by our analysis of $W^+/W^-$ ratio against the LHCb data, with up to 20\% 
deviation of the BS15 prediction versus data at $\eta_\mu>3.5$ primarily caused by a deviating behavior of the $W^-$ pseudorapidity distribution. This is
not surprising since (i) the BS15 fit has less free parameters than CT14, and (ii) the CT14 PDF fit includes also LHC (and Tevatron) $W^\pm,\,Z^0$ boson data.

Such discrepancies in the shape of the $W^-$ pseudorapidity distribution and overall normalisation versus the CMS and LHCb data are most likely due to the fact 
that the starting BS15 parametrizations were fitted to the DIS data only and have not been included into the global fit yet. Indeed, the missing NNLO/NNLL corrections 
could not substantially change the normalisation of the cross sections although may, in principle, somewhat affect the shapes of the differential distributions. 
The effect from a global fit accounting for the LHC data is expected to provide the biggest impact on the overall normalisation where the discrepancy of 
the BS model and the data are the most pronounced. This analysis should incorporate heavy quark (in particular, charm and beauty) PDFs properly as the DIS 
was not particularly sensitive to those and could not constrain them well, in distinction with the LHC data. This is an important subject for a future work that is 
needed in order to further test the reliability of the statistical approach. Finally, the global analysis of the statistical PDF model accounting for all available data 
at various energies up to LHC will verify if the number of parameters in the model should remain the same or certain modifications will be required. Clearly, the precision 
data on the absolute cross sections and individual $W^\pm$ distributions in lepton pseudorapidity at RHIC and LHC energies will be necessary for such a study.

\vskip0.5cm
{\bf Acknowledgments}~The authors would like to thank M.~Posik for helpful correspondence. R. P. was partially supported by the Swedish Research Council, contract 
number 621-2013-428, and by CONICYT grant PIA ACT1406 (Chile). E. B. is supported by CAPES and CNPq (Brazil), contract numbers 2362/13-9 and 150674/2015-5.

%%%%%%%%%%%%%%%%

\end{document}